\documentstyle[12pt,a4,epsf,amsfonts,amssymb]{article}

\def\abstract#1{\vskip 8mm 
          \begin{center}{\large Abstract}\par
\vskip 0.4cm \smallskip 
                 \begin{minipage}[c]{13cm}
                         \small #1
                 \end{minipage}
         \end{center}
}
\def\title#1{\begin{center}{\Large\bf #1}\end{center}}
\def\author#1{\vskip 5mm \begin{center}{\large #1}\end{center}}
\def\address#1{\begin{center}{\it #1}\end{center}}

\def\half{\frac{1}{2}}

\newcommand{\ben}{\begin{equation}}
\newcommand{\een}{\end{equation}}
\newcommand{\bena}{\begin{eqnarray}}
\newcommand{\eena}{\end{eqnarray}}
\newcommand{\non}{\nonumber}

\begin{document}

\begin{flushright}  
  YITP-02-44  \\
  gr-qc/0207054   
\end{flushright}
\vskip 0.3cm

\title{
    Naked Singularity and Thunderbolt 
   }

\vspace{3mm}

\author{ 
        Akihiro Ishibashi$^{\dag}$
        \footnote{\tt akihiro@yukawa.kyoto-u.ac.jp} 
        and 
        Akio Hosoya${}^{\ddag}$
        \footnote{\tt ahosoya@th.phys.titech.ac.jp} 
        } 
\vspace{2mm}  
\address{
         $^\dag$
         Yukawa Institute for Theoretical Physics,\\
         Kyoto University,\\
         Kyoto 606-8502, Japan\\
         } 
\address{
         $^\ddag$ 
         Department of Physics, \\ 
         Tokyo Institute of Technology, \\
         Oh-Okayama Meguro-ku, Tokyo 152-0033, Japan 
         }

\abstract{
  We consider quantum theoretical effects of the sudden change
  of the boundary conditions which mimics the occurrence
  of naked singularities.
  For a simple demonstration, we study a massless scalar field
  in $(1 + 1)$-dimensional Minkowski spacetime with finite spatial interval.
  We calculate the vacuum expectation value of the energy-momentum tensor
  and explicitly show that singular wave or {\em thunderbolt}
  appears along the Cauchy horizon.
  The thunderbolt possibly destroys the Cauchy horizon if its backreaction on
  the geometry is taken into account, leading to quantum restoration
  of the global hyperbolicity. 
  The result of the present work may also apply to the situation that 
  a closed string freely oscillating is traveling to a brane and 
  changes itself to an open string pinned-down by the ends satisfying 
  the Dirichlet boundary conditions on the brane. 
  }

\newpage 

\section{Introduction}

A question of whether the cosmic censorship holds
is one of the most important issues in classical general relativity,
but still remains far from being settled.
Many attempts have so far been made to prove the validity of this conjecture
and have achieved partial success under some particular
setup.
On the other hand, some of those attempts have resulted in rather showing
the counter examples,
indicating that the classical framework of general relativity
admits appearance of naked singularities~(See e.g., Review~\cite{Wald1997}).
In the present situation of research on the cosmic censorship hypothesis 
described above, it is probably more sensible
to classify naked singularities and ask whether it is a disastrous one
for physics or a practically harmless one.

In the previous work~\cite{IH1999},
as a physically sensible classification of naked singularities, 
we proposed the wave approach to probe naked singularities.
There, we studied the wave propagation in various static spacetimes
with naked singularities.
The essential point was
the uniqueness of self-adjoint extension of the time-translation
operator $A$, defined by the spatial derivatives in the wave equation.
We call the classical singularity {\em wave-regular}
if the self-adjoint extension is unique, i.e., 
there is no ambiguity in the boundary condition at the singularity.
In the wave-regular case, thus there exists a unique solution to
the wave equation for a given initial value;
the predictability holds at least for the probe fields
even in the classical framework. 
In this sense, the concept of wave-regularity is closely related to 
Clarke's idea of generalised hyperbolicity~\cite{Clarke1998}.~\footnote{%
 The question of hyperbolicity has also been discussed 
 in conical spacetimes~\cite{Vickers&Wilson2000}, and 
 in spacetimes with hypersurface singularities~\cite{Vickers&Wilson2001}.  
   }
Such a would-be singularity can be regarded as a harmless singularity. 
On the other hand, if the time evolution of an initial value is not 
unique so that the predictability for a probe wave field breaks down,
such a naked singularity is called {\em wave-singular}.

The idea of the wave probe rather than particle probe
is based on semi-classical theory, in which geometry is treated classically
but matter fields quantum mechanically, with the probe waves identified with
the mode functions for quantum fields.
One might anticipate that, once a theory of quantum gravity is established,
one would not need be afraid of the emergence of naked singularities;
a naked singularity could be resolved by replacing it with something
like a boundary of Planck energy scale, and in the neighborhood of
the resolved singular boundary, the predictability of physics would be 
restored quantum mechanically. 
But even in such a case, a remaining question will be
how such an extremely high energy boundary looks like from distant observers
who live in a low energy scale governed by (semi-) classical theories.
Then, it seems plausible that a naked `singularity' resolved
by any kind of quantum gravitational theory looks wave-regular from distant
observers, as the wave-regularity respects the predictability of low energy
(semi-) classical physics.
In this sense, naked singularities can be classified into two types:
one of which can be resolved to be wave-regular by quantum gravity,
while the other should be somehow prohibited.~\footnote{
Recently, this kind of classification of singularities has also been argued
from the stringy theoretical
view point~(See e.g., Review~\cite{Natsuume2001}).
}
Here, we would like to propose a version of (quantum-) cosmic censorship
advocating that, in semi-classical theory, wave-singular naked singularities
are prohibited to appear by quantum field theoretical effects.

In the present work, aiming at examining the above version of quantum cosmic
censorship, we shall focus our attention on the wave-singular case
in which a timelike singularity emerges
at some stage and the boundary condition is not a priori specified.
This feature may manifest itself when the singularity appears
in occasion of gravitational collapse.
A simplified version will be the case that the boundary condition suddenly
changes at some time so that the boundary becomes a naked timelike
singularity.

We shall consider such a case in the context of quantum field theory.
The time-translation operator $A$ for a field changes in time by the change of
its domain ${\cal D}(A)$ so that we naturally expect particle creations
from a vacuum state of the field.
We shall first show that this is indeed the case in the perhaps
simplest example: a massless scalar field theory in a $(1+1)$-dimensional
spacetime ${\Bbb R} \times [0,\pi]$, with the boundary condition
at $0$ and $\pi$ suddenly changing from the Neumann to the Dirichlet
boundary condition.
We shall compute the vacuum expectation value of the energy-momentum tensor
and see that it just gives the standard Casimir energy inside the domain of
dependence while it is divergent at the Cauchy horizon for the initial
hypersurface $\{0\}\times [0,\pi]$.
We will observe the thunderbolt effect~\cite{HS1993}
in the sense that there appears a delta function like singularity
which has a support only on the Cauchy horizon.
If the backreaction on the geometry is taken into account,
the thunderbolt effects will destroy the Cauchy horizon,
resulting in quantum mechanical restoration of the global hyperbolicity. 
We shall also give discussion for the general boundary conditions case 
in the appendix. 

Long ago Anderson and DeWitt~\cite{ADW} discussed a similar problem
in the context of topology change in spacetime which exhibits a splitting
of the compact universe like `trousers.'
They discussed particle creations with an infinite amount of energy
which appear an infinitely bright flash emanating from
the crotch of the trousers.
In a sense the emergence of a naked singularity is a kind of topology
change which alters the domain of definition for quantum fields. 
Not surprisingly their result, which is not fully quantitatively described, 
is qualitatively very similar to ours.

It is interesting to point out a possible application to string 
and brane theories~\cite{text:Polchinski}. 
In the brane picture of universe, an open string can 
propagate along a brane with both the ends satisfying the Dirichlet 
boundary conditions on the brane but also can pinch off from the brane 
to the bulk spacetime as a closed string. 
In that process the boundary condition at the ends changes.  
The emerged closed string will be highly excited because of the sudden 
change of the boundary condition as an analog of the phenomenon described 
in the present work.

\section{Hair of wave-singular naked singularity}

For self-containedness we briefly recapitulate the self-adjoint extension
of the time-translation operator in a static spacetime case.
The relevant mathematical materials are collected in the appendix of
Ref.~\cite{IH1999}. For thorough study of this issue,
see Ref.~\cite{RS1975}.

When probing timelike singularities in a static spacetime 
with a scalar wave $\phi$, the wave equation can take the form 
\ben
    \frac{\partial^2}{\partial t^2} \phi = - A \phi  \,,
\label{eq:KG}
\een
where $t$ is the Killing time and $A$ denotes an operator containing
only spatial coordinates and derivatives.
We will find the preliminary time-translation operator $A$
as a symmetric operator on a given Hilbert space $\cal H$
with domain $C^\infty_0$,
a set of smooth functions with compact support on $t=const.$ surface.
A trivial symmetric extension is immediately made by taking its closure.
Consider then the sets
${\cal K}_\pm := \{\phi_\pm \in {\cal H}|A^* \phi_\pm = \pm i \phi_\pm \}$,
which are called the {\em deficiency subspaces} of $A$, and
the pair of numbers $(n_+,n_-) := (\dim{\cal K}_+, \dim{\cal K}_-)$
called the {\em deficiency indices} of $A$.
If $A$ with domain ${\cal D}(A)$
is a closed symmetric operator with $n_+ = n_-$, then $A$ has self-adjoint
extensions.
Let $U$ be the partial isometries ${\cal K}_+ \rightarrow {\cal K}_-$.
Then the self-adjoint extensions ${A_E}$ can be obtained by taking
the domain as~\cite{RS1975}
\ben
  {\cal D}({A_E}) := \{ \phi_0 + \phi_+ + U \phi_+ \,
  | \, \phi_0 \in {\cal D}(A), \, \phi_+ \in {\cal K}_+ \}.
\label{def:domain}
\een

Given self-adjoint extension $A_E$, dynamics of the scalar field $\phi$
for any initial data
$(\phi(0), \partial_t \phi(0)) \in {\cal H}\times {\cal H}$
can be defined as
\bena
   \phi(t) = \cos \left( \sqrt{A_E} t \right)\; \phi(0)
            +  \frac{1}{\sqrt{A_E}} {\sin \left( \sqrt{A_E} t \right)} \;
               {\partial \over \partial t}{\phi}(0).
\label{eq:evolve}
\eena
The point is that, once a self-adjoint extension is given,
the vector $\phi(t) \in {\cal H}$ is defined not only inside the
future domain of dependence for the initial hypersurface
but also even outside the domain of dependence~\cite{Wald1980}.

In general, self-adjoint extension is not necessarily unique,
when spacetime fails to be globally hyperbolic.
Even when spacetime admits naked singularities,
if $n_+ = n_- = 0$, then $A$ has a unique self-adjoint extension
and dynamics of the probe field is uniquely determined everywhere
from the initial data by Eq.~(\ref{eq:evolve}).
This case is referred to as wave-regular. 
If $n_+ = n_- = N \neq 0$, then the partial isometry $U$ is represented by
an $N \times N$ unitary matrix $U(N)$, and $A$ has infinitely many different
self-adjoint extensions, which are in one-to-one correspondence with $U(N)$.
In this case the naked singularity is called  wave-singular. 
Each self-adjoint extension corresponds to the different boundary 
condition at the singularity and accordingly describes different
time evolution of the wave for the same initial data.
In other words, a wave-singular naked singularity has degrees of freedom
for the possible choice of the time-translation operator $A_E$.
The degrees of freedom can be interpreted as the character or the
{\em hair} of the wave-singular naked singularity,
described by the isometries $U(N)$.
For example, when a scalar probe field is considered in $(1+1)$-dimensions
of the finite spatial interval with the wave-singular
singularities at both the ends, the hair of the singularities
is $U(2)$.

It should be commented that the wave-regularity depends on the choice of
our Hilbert space $\cal H$. As emphasized in Ref.~\cite{IH1999},
the Sobolev space, which requires $L^2$ of the derivatives of $\phi$
as well as $\phi$ itself, will provide a physically sensible
Hilbert space for the singularity probe.
In the present work, however, as a simple illustration,
we shall take $L^2$-space of square integrable functions
as adopted in Ref.~\cite{HM1995}.
The analysis in the next section can be straightforwardly generalized
to the Sobolev space case. The results will somewhat change
depending on the choice of the Hilbert space.

\section{Quantum behavior of a probe field} 
\label{sec:qft}

\subsection{A simple model}

In this section, as one of the simplest cases, we shall consider
a quantum field theory of a massless scalar field in $(1+1)$-dimensional
Minkowski spacetime ${\Bbb R} \times [0,\pi] \ni (t,x)$.
As mentioned before, the hair for the boundary ``singularities'' in this case
becomes the group $U(2)$, but here we shall consider only its subgroup
$U(1) \times U(1)$ with each $U(1)$ corresponding to the arbitrariness
of the boundary conditions at $x=0, \,\pi$, respectively. 
We are concerned with quantum field theoretical effects in the situation
that the hair $U(1) \times U(1)$ suddenly changes at some time, say $t=0$,
which can be interpreted as a sudden appearance of wave-singular timelike
singularities at $x=0,\,\pi$ for $t>0$.

The scalar field $\phi$ obeys the Klein-Gordon equation~(\ref{eq:KG})
with an elliptic operator
\bena
    \quad \quad A := - \frac{\partial^2}{\partial x^2} \, .
\eena
We require the square integrability for the scalar field, i.e.,
$ \phi \in L^2(0,\pi) = \{ \phi \,| \, \int^\pi_0 dx |\phi |^2 < \infty \,\}$,
so that we can regard $A$ with ${\cal D}(A) = C^\infty_0(0,\pi)$
as a positive symmetric operator in the Hilbert space $L^2(0,\pi)$
with the inner product,
\ben
  (\phi_1,\,\phi_2) := \int^\pi_0 \! dx \; \phi_1^*(x) \phi_2(x) \,.
\een
Then, having a suitable self-adjoint extension $A_E$ of $A$,
we can obtain the solution for any initial data
$(\phi(0,x), \, \partial_t{\phi}(0,x))$ from Eq.~(\ref{eq:evolve}).

We shall focus on a particular situation that the boundary conditions
at $x=0,\,\pi$ are the Neumann boundary conditions,
$\partial \phi/\partial x = 0$, before $t=0$
but switch to the Dirichlet boundary condition, $\phi =0$ at $x=0, \,\pi$, 
after $t=0$. 
This sudden change of the boundary conditions can be viewed as a particular
case of the occurrence of wave-singular naked singularities at $x=0,\,\pi$
after $t=0$.

\subsection{Mode functions}
\label{subsec:mode-functions}

Before $t=0$ the positive frequency mode functions satisfying the Neumann
boundary condition at $x=0,\,\pi$ are given by
\bena
  f^{(in)}_n(t,x) &=& \frac{e^{-int}}{\sqrt{\pi n}} \cos nx \,,
\quad (n=1,2,\cdots)\,,
\eena
which are normalized by the Klein-Gordon inner product:
$ i\left\{(f,\; \partial_t g) - (\partial_t f,\; g)\right\}$.
Then we can expand the field $\phi(t,x)$ for $t<0$ as
\ben
   \phi(t,x) = f_0 (t,x)
               + \sum_{n=1}^{\infty}
                          \left\{
                                a^{(in)}_{n} f^{(in)}_n (t,x)
                              + {a^{(in)}_n}^\dagger{f^{(in)}_n}^*(t,x) 
                          \right\} \,,
\label{mode:<0}
\een
where
\ben
    f_0 (t,x) = \frac{1}{\sqrt\pi}(q + p\;t) \,, 
\label{zero-mode}
\een
is the zero-mode.
The expansion coefficients are required to satisfy the canonical
commutation relations
\bena
    \left[a^{(in)}_n, \; {a^{(in)}_{n'}}^\dagger \right]  
    = \delta_{nn'} \,, \quad
    \left[a^{(in)}_n, \; a^{(in)}_{n'} \right]
    =
    \left[{a^{(in)}_n}^\dagger, \; {a^{(in)}_{n'}}^\dagger \right]  
    = 0 \,,   \quad
    \left[q, \, p\right] = i \,,
\label{ccc:in-mode}
\eena
and $a^{(in)}_n$, ${a^{(in)}_n}^\dagger$ commute with $q$ and $p$
so that the canonical quantization is reproduced:
\ben
  \left[\phi(t,x) , \; \partial_t \phi (t,x')\right] = i \delta(x-x') \,.
\een

Now let us consider the time evolution of $\phi(t,x)$ into 
the future ($t>0$) region of the initial surface $t=0$, 
where the probe field $\phi(t,x)$ is supposed to satisfy 
the Dirichlet boundary condition $\phi(t,0) = \phi(t,\pi) = 0$, 
and accordingly the positive frequency mode functions are given by 
\ben
   f^{(out)}_m (t,x) = \frac{e^{-imt}}{\sqrt{\pi m}} \sin mx \,. 
\label{mode-funcs:out}
\een
In order to study the time evolution of $\phi$ into $t>0$ region,  
we need to modify the initial data at $t=0$ so as to fit
the Dirichlet boundary condition.
We shall discuss the positive frequency modes $f^{(in)}_n$ and
the zero-mode $f_0$ separately.

First, using the Fourier expansion, 
\ben
  \cos nx = \sqrt{\frac{2}{\pi}}
            \sum_{\stackrel{m=1}{(m\pm n:odd)}}^\infty
            \left(\frac{1}{m+n} + \frac{1}{m-n} \right) \; \sin mx \,,
\label{Fourier-expansion}
\een
we can express the initial data of the positive frequency modes as
\bena
   f^{(in)}_n (0,x) &=& \frac{1}{\pi} \sqrt{\frac{2}{n}}
                   \sum_{\stackrel{m=1}{(m\pm n:odd)}}^\infty
                   \left(\frac{1}{m+n} + \frac{1}{m-n} \right) \; \sin mx \,,
\label{i-data:value}
\\
  {\partial \over \partial t} f^{(in)}_n (0,x)
                    &=& - in \; f^{(in)}_n (0,x) \,.
\label{i-data:derivative}
\eena
The evolution of the mode function $f^{(in)}_n$ into $t>0$ region
is obtained by inserting the above initial data~(\ref{i-data:value})
and~(\ref{i-data:derivative}) into Eq.~(\ref{eq:evolve}),
in which ${\cal D}(A_E)$ is now spanned by the basis $\{f^{(out)}_m\}$
so as to respect the Dirichlet boundary condition.
Then we can find
\bena
  f^{(in)}_n(t,x)
             &=& \sum_{m =1}^\infty 
                 \left\{ 
                        \alpha_{nm} \; f^{(out)}_m (t,x)
                         + \beta_{nm} \; {f^{(out)}_m}^* (t,x) 
                 \right\} \,,
\label{mode-mix:positive-freq}
\eena
with the coefficients 
\bena
   \alpha_{nm} &=& \frac{2}{\pi}
               \sqrt{\frac{m}{ {}\; n{}\; }} \; \frac{1}{m - n}
               \delta_{m \pm n:odd} \;,
\quad
   \beta_{nm} = \frac{2}{\pi}
               \sqrt{\frac{m}{ {}\; n{}\; }} \; \frac{1}{m + n}
               \delta_{m \pm n:odd} \;,
\eena
where and hereafter $\delta_{m \pm n:odd}$ means  $\delta_{m \pm n:odd}=1$ 
if  $m \pm n= \mbox{an odd number}$ and $\delta_{m \pm n:odd}=0$ otherwise. 

The consistency of the canonical commutation
relations~(\ref{ccc:in-mode}) requires the unitarity relation
\ben
   \sum_{n=1}^\infty 
   \left(\alpha_{nm} \alpha^*_{nm'} - \beta^*_{nm}\beta_{nm'} \right)
   = \delta_{mm'} \,.
\een
This can be checked by using the formula
\ben
   \sum_{\stackrel{n =1}{(n:odd)}}^\infty \frac{1}{n^2} = \frac{\pi^2}{8} \,.
\een

Next, let us expand the mode which is the time evolution after $t=0$
of the zero-mode before $t=0$
in terms of the late time mode functions~(\ref{mode-funcs:out}) as
\ben
   f_0 (t, x) = \sum^\infty_{m=1}
                    \left\{
                  C_m f^{(out)}_m (t,x) + C_m^\dag {f^{(out)}_m}^* (t,x) 
                    \right\}\,.
\label{mode-mix:zero-mode}
\een
The initial data is then given by
\bena
  f_0 (0,x)&=&\sum^\infty_{m=1}(C_m+ C_m^\dag)\frac{\sin mx}{\sqrt{\pi m}}
                   = \frac{q}{\sqrt{\pi}} \,,
\\
  {\partial \over \partial t}
  f_0 (0,x)&=&\sum^\infty_{m=1}(-i m)(C_m - C_m^\dag)
                              \frac{\sin mx}{\sqrt{\pi m}}
                   = \frac{p}{\sqrt\pi} \,,
\eena
by continuity at $t=0$.
Then, by using the formula:
$(\sin mx, \sin nx) = \frac{\pi}{2}\delta_{m,n}$,
the expansion coefficients are obtained as
\bena
    C_m = \frac{2}{\pi}\frac{1}{\sqrt{m}}
                  \left( q + \frac{i}{m}p \right)\delta_{m:odd}\,.
\eena

\subsection{Particle creations}

Suppose the quantum field $\phi$ is expanded in terms of
the (out)-mode functions as
\bena
\phi (t,x) &=& \sum_m \left\{
                             a^{(out)}_{m} f^{(out)}_m (t,x)
                          + {a^{(out)}_m}^\dagger {f^{(out)}_m}^* (t,x) 
                      \right\} \,. 
\label{mode:out}
\eena
Then, from Eqs.~(\ref{mode:<0}), (\ref{mode-mix:positive-freq}),
and~(\ref{mode-mix:zero-mode}), we have the Bogoliubov transformation
\bena
  a^{(out)}_m &=& C_m
              + \sum_n \left\{
                \alpha_{nm} a^{(in)}_n  + \beta_{nm}{a^{(in)}_n}^\dagger
                       \right\} \,,
\label{Bogoliubov-1}
\\
 {a^{(out)}_m}^\dagger &=& C^\dagger_m
              + \sum_n \left\{
                       \alpha^*_{nm} {a^{(in)}_n}^\dagger
                       + \beta^*_{nm} a^{(in)}_n
                       \right\} \,.
\label{Bogoliubov-2}
\eena
This clearly exhibits the mode mixing of the positive and the negative
frequency parts, which implies particle creations 
in quantum mechanics~\cite{DeWitt1975}.

The expansion coefficients of the (out)-mode functions are now required to
satisfy the canonical commutation relations
\bena
    \left[a^{(out)}_m, \; {a^{(out)}_{m'}}^\dagger  \right]
    = \delta_{mm'} \,, \quad
    \left[a^{(out)}_m, \; a^{(out)}_{m'} \right]
    =
   \left[{a^{(out)}_m}^\dagger , \; {a^{(out)}_{m'}}^\dagger \right]
    = 0 \,,
\eena
so that $a^{(out)}_m$ and ${a^{(out)}_m}^\dagger$ are an annihilation and
a creation operator, respectively,
as well as $a^{(in)}_n$ and ${a^{(in)}_n}^\dagger$. 
Note that, for the unitarity, the contribution from the zero-mode
is indispensable. Then, defining the (in)-vacuum state as
\bena
          a^{(in)}_{n}  |\,0_{in}>= p \, |\,0_{in}>=0 \,, 
\eena
we can construct the Hilbert space of quantum states
in the Fock representation.

Now we can evaluate the expectation value of the number operator 
$N_m := {a^{(out)}_m}^\dagger a^{(out)}_{m}$ 
of the $m$-mode particle in $ |\,0_{in}>$ as 
\bena
 <0_{in}|N_m |\,0_{in}> 
      &=& <0_{in}| C^\dagger_m C_m |\, 0_{in}> \delta_{m:odd}
       + \sum_n |\beta_{mn}|^2
\non \\
    &=& \frac{4}{\pi^2}\left(
                     \frac{<0_{in}|q^2 |\, 0_{in}>}{m} - \frac{1}{m^2}
                       \right) \delta_{m:odd}
      + \sum_n \frac{m}{n} \; \frac{1}{(m + n)^2} \,.
\eena
This is finite, but the total number of the created particles
$\sum_{m=1}^\infty  <0_{in}|N_m |\,0_{in}> $ is logarithmically divergent,
so that the Bogoliubov transformation  is not
unitarily implementable~\cite{Kodama1980}.
This means the final Fock space is not equivalent to but completely different
from the initial Fock space:
\ben
    < 0_{out}|\,0_{in}>=0 \,. 
\een
This also suggests that the physics completely changes
by the emergence of wave-singular naked singularities.
We can see in the subsequent section that this is indeed the case from
the study of the vacuum expectation value of the energy-momentum tensor
before and after the incident at $t=0$.

\subsection{Thunderbolts}
\label{subsec:thunderbolts}

Let us see the physical effects of the particle creations
by the sudden change of the boundary conditions,
computing the vacuum expectation value of the energy-momentum tensor.
It is convenient to introduce the advanced and the retarded time
$z_\pm :=t \pm x$, respectively.
Then, from Eqs.~(\ref{mode:<0}) and~(\ref{zero-mode}),
we can compute the renormalized energy-momentum tensor in $ t<0$ region as
\bena
<0_{in}|\,T_{\pm \pm} \, |\,0_{in}>_{t<0}
       &=& <0_{in}| \, (\partial_\pm \phi)^2 \, |\,0_{in}>
\non \\
     &=& \frac{1}{8}\sum_{n=1}^\infty n \;
         + <0_{in}|\,(\partial_\pm f_0)^2 \,|\,0_{in}>   
\non \\
       &=& - \frac{1}{96}  \,,
\eena
where we have adopted the $\zeta$-function regularization~\cite{BD}. 
The right hand side is the Casimir energy due to the finiteness of
the spatial section.
 
\begin{figure}[h]
  \centerline{\epsfxsize = 9.0cm \epsfbox{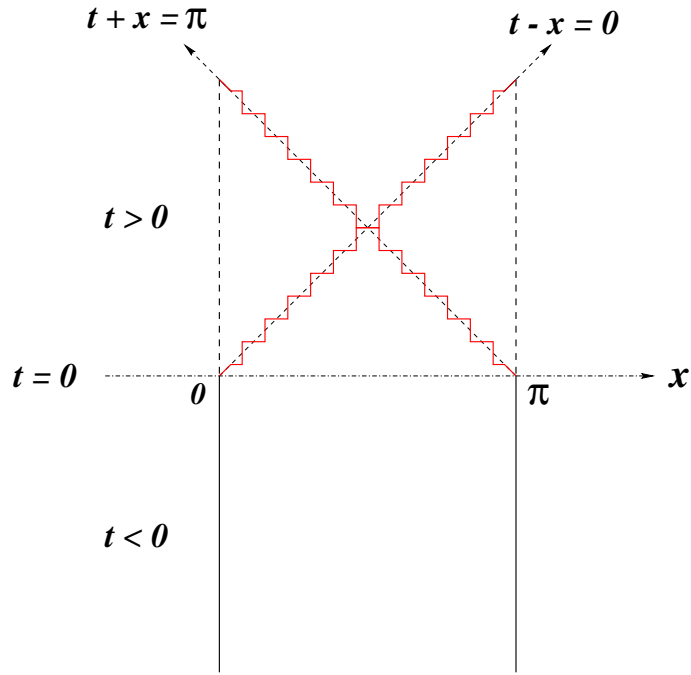}}
\vspace{3mm}
  \begin{center}
  \begin{minipage}[c]{13cm}
   \caption{ 
           The solid and dashed vertical lines denote the boundaries 
           at $x=0,\,\pi$, where the Neumann and the Dirichlet boundary 
           conditions are imposed, respectively. 
           The dashed vertical lines mimic 
           the emerging naked singularities.  
           Thunderbolts, denoted by zigzag lines, appear 
           along the null lines, $z_+ = \pi$ and $z_- = 0$. 
           } 
         \protect \label{fig:Thunderbolt}
  \end{minipage}
  \end{center}
\end{figure}

On the other hand, from Eq.~(\ref{mode:out}), and the Bogoliubov
transformation~(\ref{Bogoliubov-1}) and (\ref{Bogoliubov-2}),
we can have in $t>0$ region,
\bena
  <0_{in}|\,T_{\pm \pm} \, |\,0_{in} >_{t>0}
      &=& - \frac{1}{96}
        + \left(\frac{1}{\pi}<0_{in}|\,q^2\,|\, 0_{in}>
        + \half\sum_{n=1}^{\infty} \frac{1}{n} \right)
          \left\{ \delta(z_\pm) - \delta(z_\pm - \pi) \right\}^2 \,,
\label{Tmunu}
\eena
with the aid of the formula
\bena
   \sum_{\stackrel{l = -\infty}{(l:odd)}}^{\infty} \frac{e^{-ilz}}{l}
    = - \pi i \left\{ \Theta(z) - \Theta(z-\pi) - \frac{1}{2} \right\}  \,,
\quad
      \sum_{\stackrel{l = -\infty}{(l:odd)}}^{\infty} e^{-ilz}
    = \pi \left\{ \delta(z) - \delta(z-\pi) \right\} \,. 
\label{formula:step-delta-function}
\eena
The first term in the right hand side of Eq.~(\ref{Tmunu}) 
is the Casimir energy, 
which is present also in the $t<0$ case. From the second term, it turns out 
that the energy density diverges along the null
boundaries $z_+ = \pi$ and $z_- = 0$, which form the Cauchy horizon for
the $t=0$ initial surface as illustrated in Fig.~\ref{fig:Thunderbolt}.
Besides the logarithmically divergent factor,
the square of the delta function implies that not only
the energy density but also the total energy will be unbounded.
In other words, classical observers passing through the Cauchy horizon
will see the {\em thunderbolts}.

\section{Summary and Discussion}

We have studied quantum effects of the sudden change of the boundary
conditions which mimics the occurrence of wave-singular
naked singularities.
To be specific, we have dealt with the massless scalar field
in $(1 + 1)$-dimensional Minkowski spacetime with finite spatial interval.
We have shown that the particle creations occur due to
the sudden change of the boundary condition
from the Neumann to the Dirichlet condition at $t=0$.
Quantum particle creation is known to happen when the time-translation
operator, or the Hamiltonian is time dependent.
As we have emphasized, an operator of the spatial part of the wave equation
is defined as a pair of the action $A$ and
its domain of definition ${\cal D}(A)$. The sudden change of
the boundary condition is nothing but the change of the domain ${\cal D}(A)$
so that in this sense the time-translation operator is time dependent.
The behavior of the Bogoliubov coefficients in this model implies
that the unitarity implementability breaks down, and 
the Fock space for $t>0$ is not equivalent to the one for $t<0$. 
Furthermore, having computed the vacuum expectation value of 
the energy-momentum tensor, we explicitly showed that, 
the singular null waves or the thunderbolts appear along the null 
lines emanating from the edges of the emerging singularities, 
while the energy-momentum tensor is the Casimir energy inside the future 
domain of dependence of the initial hypersurface.

Our result suggests that the Cauchy horizon associated with
wave-singular naked singularities will be destroyed by the thunderbolts
and the global hyperbolicity will be restored by the quantum effects.
Thus this will be a support for quantum cosmic censorship
stating that wave-regular naked singularities could appear
but wave-singular ones are not admitted to emerge
when quantum theoretical effects are taken into account.

Although the present analysis is restricted to the simplest
case such that the boundary conditions change from the Neumann condition
to the Dirichlet condition, it is possible to explicitly work out
the general boundary condition case. 
Since $U(N)$ parameters which appear in general boundary condition
characterize the hair of a wave-singular naked singularity,
the calculation of the general boundary conditions case will tell us
the relation between the strength of thunderbolt
and a hair of the wave-singular naked singularity. 
A brief discussion of the general boundary condition case is given 
in the appendix.

In the present work we have examined only the $(1+1)$-dimensional model,
for explicit demonstration.
The results might be applicable even to the 4-dimensional spacetime case,
following the discussion by Anderson and DeWitt~\cite{ADW}.
The point is that the appearance of the thunderbolts is due to the sudden
change of the domain ${\cal D}(A)$ of $A$ at the initial surface.
The discontinuous change of the basis functions $\{f_n\}$ of ${\cal D}(A)$
implies a delta-function like behavior of the gradient $\nabla f_n$
at the initial surface.
Since the energy-momentum tensor is bilinear in $\nabla f_n$,
the square of the delta-function, like the second term of Eq.~(\ref{Tmunu}),
are naturally expected to appear in a $4$-dimensional setup.

So far there have appeared many works on quantum effects
around a Cauchy horizon, showing the divergence of the expectation value 
of the energy-momentum tensor along or near 
the horizon~(See e.g., Review~\cite{QFTatCH} and references therein). 
The particle creation occurs as a result of the spacetime dynamics
just {\em before} the instant of the naked singularity formation
by gravitational collapse (See e.g., Ref.~\cite{HINSTV}), 
but is not directly related to the nature of the singularity itself.
On the other hand, in the present simple analysis, 
no dynamics of the spacetime metric has been considered. 
The appearance of the thunderbolts may be interpreted as a result of 
the wave scattering off at $t=0+$ by the boundary singularities.  
We can extend our present analysis to a gravitational collapse model in 
$(3+1)$-dimensions. Then as commented above, we expect a quantum effect, 
leading to the thunderbolt. This is a new phenomenon, 
because it is not merely a consequence of dynamics which creates 
naked singularities but directly reflects the character of 
the wave-singular naked singularities.

\section*{Acknowledgments}
A.I. thank Professor H. Kodama for useful comments.  
This work is supported in part by the Japan Society 
for the Promotion of Science (A.I.) and the Ministry of Education, 
Science and Culture of Japan under grant no. 09640341 (A.H.).

\section*{Appendix} 

Here we shall consider the general boundary condition case in which   
the boundary conditions at $x = 0, \, \pi$ are given as 
$b \, \partial_x\phi = \phi $ before the incident $t=0$ 
but suddenly changes to $a \,\partial_x \phi = \phi $ after $t=0$, 
with $a,b $ being constants.   
Note that when $a,b=0$, the field satisfies the Dirichlet boundary condition 
and when $a,b \rightarrow \infty$, the Neumann boundary condition.

The positive frequency (in)- and (out)- mode functions 
are respectively given by 
\bena 
   f^{(out)}_m (t,x) &=& \frac{e^{-imt}}{\sqrt{\pi m}} 
                    \frac{1}{\sqrt{1 + a^2m^2}} 
                    \left( \sin mx + a m \cos mx \right) \,,
                                        \quad \mbox{in $t>0$} \,, 
\\
   f^{(in)}_n (t,x) &=& \frac{e^{-int}}{\sqrt{\pi n}} 
                    \frac{1}{\sqrt{1 + b^2n^2}} 
                    \left( \sin nx + b n \cos nx \right) \,,
                                        \quad \mbox{in $t <0$} \,, 
\eena 
so that $ b\,\partial_x {f^{(in)}_n} = f^{(in)}_n $ 
and $ a\,\partial_x {f^{(out)}_m} = f^{(out)}_m $ at $x=0,\, \pi$. 
Besides these positive frequency modes, 
when $a$ or $b\rightarrow \infty$, there exist zero-modes.  
But here we do not treat the contributions from zero-modes, 
just for simplicity.

Then, following the same step in Section~\ref{subsec:mode-functions}, 
we can find the Bogoliubov coefficients:  
\bena
\alpha_{nn} &=& \frac{1 + abn^2}{\sqrt{1+a^2n^2}\sqrt{1+b^2n^2}}
                \delta_{m,n} \,, 
\\
\alpha_{mn} &=& \frac{2}{\pi}(b-a)
                \frac{\sqrt{mn}}{\sqrt{1+a^2m^2}\sqrt{1+b^2n^2}}\frac{1}{m-n}
                \delta_{m \pm n:odd} \,, 
\\ 
\beta_{mn} &=& \frac{2}{\pi}(b-a)
                \frac{\sqrt{mn}}{\sqrt{1+a^2m^2}\sqrt{1+b^2n^2}}\frac{1}{m+n} 
                \delta_{m \pm n:odd} \,.  
\label{Bogol-coeff:beta}
\eena 
From this we can immediately see that, 
if $ab \neq 0$, then $\sum_{n=1}^{\infty} \sum_{m=1}^{\infty} |\beta_{nm}|^2$ 
is finite, while if $ab = 0 \, (a \neq b )$, 
then it is logarithmically divergent, hence in this case 
the Fock space of the (in)-states and that of the (out)-states 
are unitarily inequivalent.

The vacuum expectation value of the energy-momentum tensor can be computed 
in the same way in Section~\ref{subsec:thunderbolts}. 
The result can be written down in terms of Lerch's transcendent, 
which is rather complicated and not so illuminating for drawing 
the physical consequence.

\end{document}